# Assignment of collision-induced four-level double-resonance transitions in the $3\nu_3 \leftarrow \nu_3$ spectral region of methane

Kevin K. Lehmann[1,+], Isak Silander[2], Adrian Hjältén[2], Michael Rey[3], and Aleksandra Foltynowicz[2,*]

[1]Departments of Chemistry and Physics, University of Virginia, Charlottesville, Virginia 22904, USA
[2]Department of Physics, Umeå University, 901 87 Umeå, Sweden
[3]Groupe de Spectrométrie Moléculaire et Atmosphérique, UMR CNRS 7331,BP 1039, F-51687 Reims Cedex 2, France

[+]kl6c@virginia.edu, *aleksandra.foltynowicz@umu.se,

**Abstract**

Optical-optical double-resonance (OODR) spectroscopy using a narrow-linewidth pump and a frequency comb probe has previously been used to measure and assign sub-Doppler transitions in the $3\nu_3 \leftarrow \nu_3$ spectral region [J. Chem. Phys. **161**, 124311 (2024)] when pumping from the J = (7, $A_2$) ground state. Doppler-broadened double-resonance transitions were also observed in those OODR spectra. In this paper, 68 of these Doppler-broadened transitions are assigned to four-level double-resonance transitions involving collisional transfer from the pumped $A_1$ symmetry state to other $A_1$ and $A_2$ symmetry (I = 2 meta nuclear spin) levels of the $\nu_3$ fundamental state. Assignments are made using combination differences and comparison with the term values and intensities of lines predicted by a new effective Hamiltonian, the accuracy of which has been validated by the sub-Doppler transitions.

## 1. Introduction

Methane has long been of keen interest to molecular spectroscopists as the paradigm of a spherical top molecule with $T_d$ point group symmetry.[1,2] Paradoxically, the high symmetry and resulting degeneracies makes the spectroscopy more rather than less complicated.[3,4] All four of the normal modes of vibration are strongly coupled by a series of Fermi, Darling-Dennison, and Coriolis resonances, leading to extensive mixing of the zero-order harmonic oscillator-rigid rotor states that share the same polyad number,[5] defined as $P = 2\nu_1+\nu_2+2\nu_3+\nu_4$ with $\nu_i$ the total number of quanta on mode i.

Methane plays an important role in multiple environments, including the atmosphere of the Earth[6] and planets of our Solar system,[7-10] exoplanets,[11-16] brown dwarf stars,[17] as well as in combustion.[18-20] Remote sensing of methane spectra can be used to probe such environments, but this requires the ability to accurately predict the observed spectra.[21, 22] This is particularly challenging in hot environments, such as hot-Jupiter exoplanets and combustion. Recent advances in computational methods[21, 23-31] have been used to predict such spectra using massive lists of spectroscopic transitions including many hot bands and high rotational states. These lists have greatly aided the assignment of experimental spectra, which in turn provide tests of the accuracy of the predicted spectra and can be used to further improve the models.[32]

Despite the predictions, assignments of hot methane spectra are challenging due to the spectral complexity and line overlap at high (compared to ambient) temperatures.[33] Optical-optical double-resonance (OODR) spectroscopy provides a powerful tool to detect transitions from selected excited rovibrational states and to determine the rotational quantum number and symmetry of the observed final states.[34-40] We have previously introduced the use of a frequency comb as the probe radiation source in OODR spectroscopy, allowing the parallel detection of sub-Doppler transitions over a few THz of bandwidth[35, 36] with cavity-enhanced absorption sensitivity.[34] So far, we used the technique to measure and assign double-resonance transitions of methane arising from individual rotational states of the $\nu_3$ asymmetric stretching vibrational fundamental to the states near the $3\nu_3$ states (P6 polyad).[34, 36, 41] These initial states were populated by excitation of the ground state using a 3.3 µm continuous-wave optical parametric oscillator that was Lamb-Dip locked to the center of the selected pump transition. The $3\nu_3 \leftarrow \nu_3$ Doppler-free transitions, whose lower state was the upper state of the pump transition, were probed by a frequency comb tunable in the ~1.7 µm spectral region. These transitions are known as ladder-type 3-level double resonance (3LDR) transitions, and their initial state is known from the assignment of the pump transition and the symmetry species of the final state defined by the methane spectroscopy selection rules ( $A_1$ <-> $A_2$; $F_1$ <-> $F_2$, E <->E ). In our most recent work,[41] the pumped states had rotational quantum numbers J(R) = 6(7), 7(7), and 8(7), all of $A_1$ rovibrational symmetry. Here, J is the total angular momentum (excluding nuclear spin) quantum number and R is the quantum number of the rotational angular momentum. We detected 118 3LDR transitions, reaching 84 unique final states with J between 5-9. These upper states were assigned by combination differences and comparison with theoretical prediction of the wavenumber and intensity of the transitions starting from the pumped rotational levels of the $\nu_3$ fundamental. Three different theoretical spectra were used; TheoReTS/HITEMP,[30, 42] ExoMol,[27] and new effective Hamiltonian[25] predictions, where the latter were found to be the most accurate with a mean offset of -0.16 cm$^{-1}$ and a residual standard deviation of 0.18 cm$^{-1}$.

In addition to the Doppler-free transitions, we observed many Doppler-broadened (DB) double-resonance absorption transitions without a sub-Doppler component. In our prior publication,[41] we suggested that these transitions had lower states that were populated by collisions from the pumped $\nu_3$ level; they are known as four-level double-resonance (4LDR) transitions as the pump and probe transitions have no state in common. In this work, we propose assignments for many of these 4LDR transitions using combination differences and comparison to the new effective Hamiltonian predictions.[25]

## 2. Experimental

The instrument was previously described[41] and the present results are obtained by reexamination of the previously reported spectra. Briefly, we used a CW optical parametric oscillator (TOPO, Toptica) tunable around 3.3 µm to pump the P(7,$A_2$), Q(7,$A_2$), and R(7,$A_2$) transitions in the $\nu_3$ fundamental (C-H asymmetric stretch) band and a 250 MHz near-IR frequency comb tunable around 1.7 µm to probe transitions from the states populated by the pump, both directly and after collisional relaxation. The pump was Lamb-dip locked to the selected transition using a reference cell. The sample gas of high purity methane at a pressure of 200 mTorr (26.7 Pa) was held in a cavity with a finesse of ~1000 for the comb, and transmitting the pump. The cavity had a length of 60 cm and the $TEM_{00}$ mode of the probe field had a beam radius at the focus of 0.50 mm and a Rayleigh range of 45.8 cm. The pump beam was shaped to have the same focal point and Rayleigh range as the probe, and thus had a beam radius at focus of $w_p$ = 0.70 mm. Probe spectra were measured using a Fourier transform spectrometer with auto-balanced detection at different comb repetition rates and interleaved to produce absorption spectra with a sample point spacing of ~2 MHz in the optical domain. Probe spectra with pump on and pump off were ratioed to largely remove the transitions arising from thermally populated $CH_4$ levels. Further details of the data processing can be found in the original publications.[34, 41]

## 3. Line detection and fitting

Previously, we focused on the 3LDR sub-Doppler probe transitions which had a half-width at half maximum (HWHM) of ~10 MHz, dominated by the power broadening produced by the pump transition. This time, we searched for the DB 4LDR transitions in the same spectra. For this, the normalized probe spectra were run through a peak-finding routine similar to that used earlier[41] but with parameters optimized to find peaks with width similar to the expected thermal Doppler width. First, we filtered the spectra in two different ways: i) using a bandpass filter of 20 MHz to 20 GHz and ii) by convolving the spectrum with a Lorentzian dispersion line shape with a HWHM of 240 MHz and taking a derivative of this convolution. We multiplied the two processed spectra, which yielded a spectrum with peaks at the positions of DB lines, with higher signal-to-noise ratio (SNR) than in the original spectrum. We searched for these peaks using the MATLAB *findpeaks* function. Detected lines with an SNR larger than two in the filtered spectrum were manually sorted into DB peaks, sub-Doppler peaks, and false detections. The sub-Doppler peaks were also detected by this peak-finding routine because they sit on a Doppler-broadened background, attributed to absorption of molecules that had undergone elastic collisions before leaving the probe volume. A narrow portion of the P(7,$A_2$) – pumped spectrum is shown in Fig. 1, displaying one 3LDR transition and two 4LDR transitions. The sidebands on the 3LDR transition come from the frequency modulation of the pump beam used for the Lamb-dip locking.

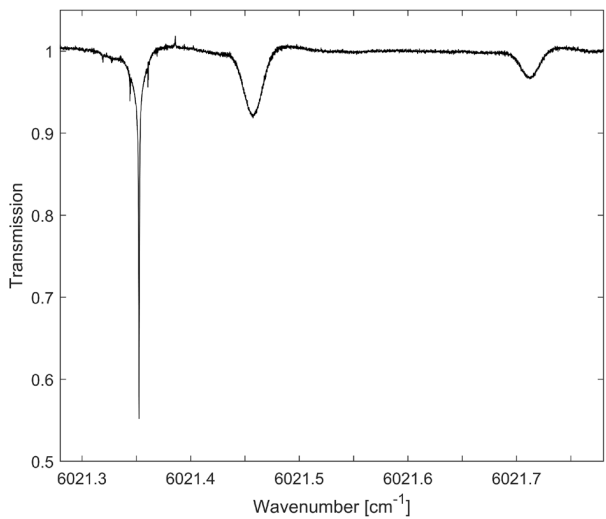

Fig. 1. Part of the normalized P(7,$A_2$)-pumped spectrum with a 3LDR line around 6021.35 cm$^{-1}$ and 4LDR lines around 6021.45 cm$^{-1}$ and 6021.7 cm$^{-1}$.

The cavity transmission in a ±1 GHz range around each identified DB peak was then fit using a cavity-enhanced Gaussian line shape model that included the shift of cavity mode resonance due to the resonant dispersion of the methane gas.[43] The cavity finesse values were taken from the previous work.[41] The fitting parameters were the center frequency and integrated absorption coefficient, as well as the Doppler-broadened width (restricted to the range of 200 to 400 MHz), and the comb-cavity phase offset (restricted to ±1 mrad). If the fit did not converge or if any of the parameters hit a limit of the fit the following procedure was applied: first, the fitting range was restricted to ±400 MHz. Next, if this did not make the fit converge, the range was restored to ±1 GHz, but a background offset was introduced in the fit. Finally, a first order polynomial baseline was added to the fit with the symmetric fitting range. For lines spectrally overlapping with neighboring features, an asymmetric fitting range of ±200 MHz to ±1 GHz was used. We estimated the total line center uncertainties as a combination of fit uncertainties, and two contributions stemming from the pressure shift and the calibration of the FTS reference laser wavelength, as described previously.[41]

Overall, we identified 318 4LDR transitions between 5905-6099 cm$^{-1}$ when pumping the P(7,$A_2$) transition; 263 between 5866-6086 cm$^{-1}$ when pumping the Q(7,$A_2$) transition, and 86 between 5826-5979 cm$^{-1}$ when pumping the R(7,$A_2$) transition. The spectral range probed was shifted between pumping the P(7,$A_2$), Q(7,$A_2$), and R(7,$A_2$) lines to offset the shifts in wavenumber of these pump transitions, so that final state term values for the 3LDR transitions in each case overlapped as that maximized the use of combination differences to confirm rotational assignments. A 4LDR transition starting from the same collisionally populated level will have a wavenumber independent of the pump transition. In the overlapping regions of the three spectra, most of the strong DB transitions in one spectrum appeared in the other, which hence supports their assignment as 4LDR transitions. For each DB transition observed in more than one spectrum, we calculated the mean of the retrieved center frequency and integrated absorption coefficient, weighted by the inverse of the square uncertainties. Some of the 4LDR transitions were found at the same wavenumber as 3LDR transitions in other J = (7,A$_2$) – pumped spectra; in those cases, we replaced the center frequencies by those reported in[41, 44]. Figure 2 shows a stick spectrum of all identified lines, compared to predictions from the effective Hamiltonian[25]. In the Supplementary material, we give a table with the complete set of the 375 detected 4LDR lines with their fit center wavenumbers and integrated absorptions. The uncertainties in the latter include a 4% contribution from the cavity finesse, in addition to the fit uncertainties.

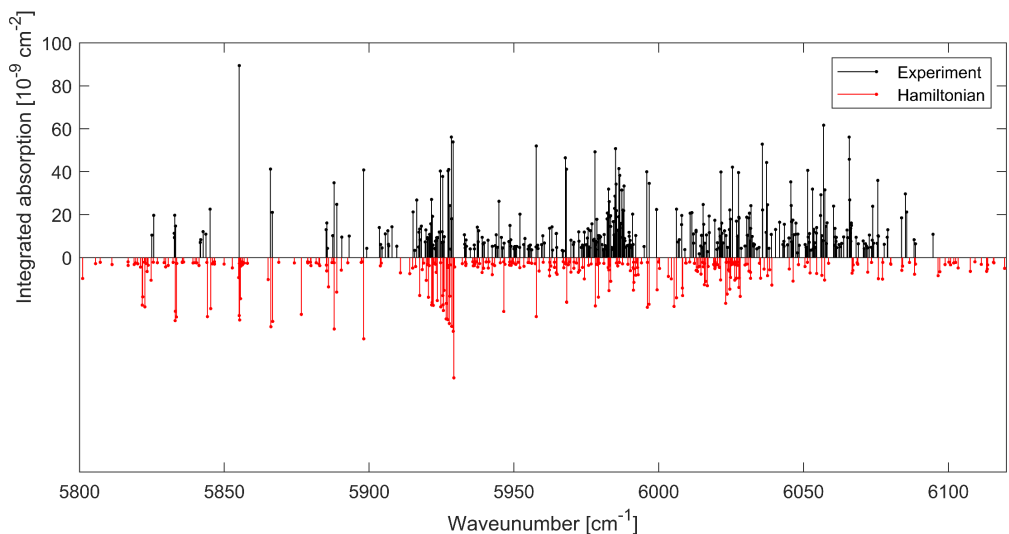

Fig. 2. The integrated absorption of all detected lines (black) compared to Einstein A coefficients from the Hamiltonian predictions (red). The predicted Einstein A-coefficients are scaled to match the experiment and plotted inverted for clarity.

## 4. Collisional transfer

CH$_4$ gas has a reported self-diffusion constant of D = 866 cm$^2$/sec at 200 mTorr.[45] Based on this, a mean free path of 0.164 mm and a hard-sphere collision rate of 3.86/µsec is calculated. Taking the inelastic rate equal to this rate, we predict a pressure broadening coefficient (HWHM) of 0.078 cm$^{-1}$/atm, in good accord with the self-broadening values listed in HITRAN which range from 0.081 – 0.074 cm$^{-1}$/atm from J" =0 to 8.[46] The time required for diffusion in the transverse plane perpendicular to the optical axis a distance equal to the radius of the pump beam is calculated as $t_D$ = w$_p^2$/4D = 1.41 µs, and thus about 5 times the mean time between inelastic collisions. Based upon pressure broadening coefficient of the Lamb dip in the P(7) transitions of the ν$_3$ band, 0.41(2) cm$^{-1}$/atm,[47] the elastic collision rate is known to be about 6 times larger than the inelastic rate. The elastic rate is dominated by small angle, large impact parameter scattering.[48]

If we assume that at t = 0 we pump molecules with a spatial distribution equal to the intensity of the pump beam, then in the diffusion dominated limit, solving the diffusion equation predicts that the distribution of vibrationally excited molecules will be given by

$$N(r,t) = \frac{1}{4\pi D(t+t_D)} exp\left(-\frac{r^2}{4D(t+t_D)} - \gamma_v t\right) \tag{1}$$

where $\gamma_v$ is the rate of vibrational relaxation and *r* is the distance from the optic axis. If we take the rate of production of vibrationally excited molecules per unit distance as $\alpha_p P_p/h\nu$, where $\alpha_p$ is the saturated absorption coefficient of the pump laser which has power $P_p$, integration of N(0,t) from t = 0 to infinity predicts a steady state on-axis (r =0) density of vibrationally excited molecules

$$\rho(r) = \frac{\alpha_p P_p}{4\pi D h\nu} \Gamma(0,\gamma_v t_D)e^{-\gamma_v t_D} => \frac{\gamma_e}{\gamma_v t_D} \text{ as } \gamma_v t_D \ll 1 \tag{2}$$

where $\Gamma(0,\gamma_v t_D)$ is the incomplete Gamma function and $\gamma_e = 0.577 \ldots$ is Euler's constant.

C.B. Moore and coworkers published a series of papers where they pumped individual transitions of CH$_4$ and monitored IR emission in the regions of the $\nu_3$ (3.3 µm) and $\nu_4$ (7.6 µm) IR allowed fundamentals.[49-56] In the earliest work, Yardley and Moore[53-56] excited CH$_4$ with a amplitude modulated 3.39 µm HeNe laser which overlaps the P(7) transitions of the $\nu_3$ fundamental. Later, Hess and Moore[49-51] used a pulsed OPO to excite Q branch lines in the $\nu_3$ fundamental and they monitored the CH$_4$ IR emission through a filter that allowed transmission of the R(J > 4) lines. The rise of this emission was instrument limited (100 ns), despite requiring changing the vibrational angular momentum projection quantum number for R branch emission. The fall of the emission was fit to a single exponential with decay time times pressure given by 3.9(6) µs Torr, which they interpreted as 50 times longer than the mean time between hard-sphere collisions times pressure (78 ns Torr). Surprisingly, the rise of $\nu_4$ emission was somewhat faster, 2.8(6) µs Torr and had a very long decay, consistent with the very long V → R,T relaxation time of the $\nu_4$ and $\nu_2$ states, which reach relative equilibrium faster than the decay of $\nu_3$. In order to distinguish emission from the $\nu_4$ fundamental state and from the states $\nu_2 + \nu_4$ and $2\nu_4$, they used a CH$_4$ gas cell between excitation region and the IR detector to block >95% of the $\nu_4$ fundamental emission. This gave a weak biexponential emission signal with rise of 1.6(8) µs Torr and fall of 4.0(1.5) µs Torr. The decay time was interpreted as the lifetime of the $\nu_3$, which feeds the $\nu_2 + \nu_4$ and $2\nu_4$ states and the rise time of the signal the lifetime of the $\nu_2 + \nu_4$ and $2\nu_4$ states to decay into the $\nu_2$ and $\nu_4$ fundamental levels, likely mostly by collisions that transferred a bending quantum to another CH$_4$ molecule, which is a nearly isothermal neutral process. They considered the process where the $\nu_3$ fundamental relaxes to the $\nu_1$ fundamental (releasing 103 cm$^{-1}$) but concluded from the single exponential decay of $\nu_3$ emission that this process must either be very fast (no more than a few hard-sphere collision times) or very inefficient compared to the relaxation into the bending combination states. In a subsequent paper, Hess, Kung, and Moore[49] used a more powerful and narrower band OPO to excite $\nu_3$ and also states $2\nu_3$, $\nu_3+ \nu_4$, and $\nu_1+ \nu_4$. The decay from the $\nu_3$ fundamental excitation agreed with the earlier result but the rise of the $\nu_4$ emission now matched the $\nu_3$ decay. They determined the time for the $\nu_3+\nu_4 \leftrightarrow \nu_1+ \nu_4$ transfer to be 0.8(2) µs/Torr and suggested that the same rate can be used to estimate the $\nu_3 \leftrightarrow \nu_1$ transfer. However, the pumped levels of $\nu_3+\nu_4$ and $\nu_1+\nu_4$ are both of $F_2$ vibrational symmetry and mixed by anharmonic interactions, while $\nu_1$ and $\nu_3$ are not, so this may not provide an accurate estimate of $\nu_3$ and $\nu_1$ interconversion.

Klaassen et al.[57] used an $H_2$ gas Raman-shifted pulsed Ti:sapphire laser to excite a series of transitions in the $2\nu_3$ and the $\nu_3+\nu_4$ bands and monitored the time dependent changes of transmission using a 3.3 μm CW diode laser to monitor both 3 and 4 level V and Λ type double resonance transitions using the $\nu_3$ fundamental for the V type and $2\nu_3 – \nu_3$ and $\nu_3+ \nu_4 –\nu_4$ difference frequency bands for the Λ type. 3LDR transitions were used to monitor the total rates of state relaxation from (for Λ type) or to (for V type) the state involved in the pump transitions. 4LDR transitions were used to observe the time dependent population in other states and kinetic models used to determine state-to-state rates. They determined that the total rate of state relaxation was comparable to, but systematically less than, the pressure broadening rate of the pump transitions, indicating the importance of $M_J$ (angular momentum reorientation) changing and pure dephasing collisions. The 3LDR transitions showed substantial differences when parallel vs perpendicular polarization of the pump and probe fields was used, demonstrating the importance of $M_J$ changing collisions. They determined that when pumping $A_1$ or $A_2$ symmetry states, population only flowed into other $A_1$ and $A_2$ symmetry states, with a propensity rule that parity conserving rates were larger than parity changing rates. Recall that $CH_4$ has three different nuclear spin states with total nuclear spin of I = 2, 1, and 0 and that these are associated with states of rovibrational symmetry $A_{1,2}$, $F_{1,2}$, and E respectively. Collisional interconversion between states of different nuclear spin is known to be extremely slow,[58] but near resonant vibrational excitation transfer (swap) in $CH_4$ pair collisions could populate ro-vibrational excited states of $F_{1,2}$, and E symmetries. Klaassen *et al*. determined that the rate of $\nu_4$ vibrational swap between pairs of $CH_4$ molecules was 0.35(5) μs$^{-1}$ Torr$^{-1}$, close to the rate of 0.47 μs$^{-1}$ Torr$^{-1}$ predicted by a Förster resonant transfer model.[59] In contrast, while they reported that the predicted Förster resonant transfer rate for $\nu_3$ is 0.38 μs$^{-1}$ Torr$^{-1}$, their results indicate that the rate of this process must be "substantially smaller".

At 200 mTorr, the time for diffusion out of the probe volume in our experimental geometry is $\tau_D = 1.41$ μs, while the predicted lifetime for the vibrational relaxation of $\nu_3$, based upon the Moore results is 19(4) μs, i.e. $\gamma_v \tau_D = 0.074$. Hess *et al*'s lifetime estimate[51] for relaxation time between $\nu_3$ into $\nu_1$ is 4 μs, which we argue is likely an underestimate, and thus considerably longer than time for diffusion to remove the pumped molecules from the probe volume. If we take the $\nu_3$ vibrational swap rate calculated by the Förster model, which Klaassen *et al*.[59] report is substantially too fast, we get a lifetime for $\nu_3$ into $\nu_1$ as 15.2 μs, also much longer than the diffusion time out of the probe volume. Based upon these considerations, we can conclude that in our experiment, the 4LDR transitions should be dominated by transitions starting from other $A_1$ and $A_2$ symmetry states of the $\nu_3$ fundamental.

## 5. Line assignment

Some of the Doppler-broadened 4LDR transitions were found at the same wavenumber as 3LDR transitions in other J = (7,$A_2$) – pumped spectra[41] as well as in R(0,$A_1$) – pumped spectra (unpublished), but not at the wavenumbers of 3LDR transitions previously observed when pumping the $2F_2$ state.[34] This is consistent with the above mentioned rates. We looked for predicted strong transitions in our spectral window from $A_1$ and $A_2$ symmetry states of $\nu_1$ to the $P_6$ states that have been assigned and thus have precisely known wavenumbers. None of these transitions were observed in our 4LDR line list, so we can discount collisional relaxation from $\nu_3$ -> $\nu_1$ as a significant relaxation pathway.

We therefore focused on assignments on transitions from $A_1$ and $A_2$ symmetry states of $\nu_3$, shown in Fig. 3. For each value of J and $M_J$ there are (2J+1) different rotational states that can be written as linear combination of symmetric top rotational states. For a rigid spherical top, these would all be degenerate. However, centrifugal distortion and ro-vibrational interactions partially split these states into sets of states that transform as irreducible representations of the $T_d$ group. Excitation of ground level for J with symmetries $A_1$ or $A_2$ will produce a level of symmetries $A_2$ or $A_1$ respectively with J – 1 (P branch), J (Q branch) and J+1 (R branch).

Unlike for the case of transitions from $A_1$ vibrational states, vibrational transitions starting in $F_2$ symmetry vibrational states (such as the $\nu_3$ fundamental) have fully allowed P, Q, and R branch transitions to individual final quantum states,[38] and thus combination differences can be used for assignment. We also used the theoretical line list based on a new effective Hamiltonian[25] which - based on comparison with the 3LDR spectra - predicts spectral lines in this region with a mean offset of 0.16 cm$^{-1}$ and standard deviation of 0.18 cm$^{-1}$ for the J=5-9 states,[41] and -0.04 and 0.13 cm$^{-1}$ respectively for transitions with J = 2-4.[60] Further, the ratio of observed and predicted line intensities, after an overall scale factor, were predicted with a standard deviation of 0.32[41], which can be used to screen potential combination difference pairs.

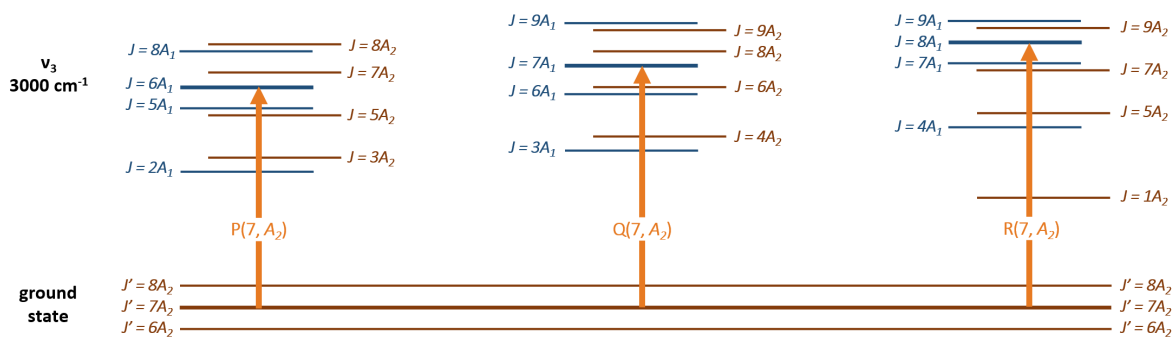

Fig. 3. Schematic illustration of the $CH_4$ energy levels relevant for the 4LDR transitions. The orange arrows show the three pump transitions starting from the J' = (7, $A_2$) state reaching J = (6, $A_1$), (7, $A_1$) and (8, $A_1$) states in the $\nu_3$ band (indicated in bold). The $A_1$ and $A_2$ symmetry states in the $\nu_3$ band that can be populated by collisional transfer are also indicated. The collisional conservation of nuclear spin states[61] restrict relaxation to a subset of methane ro-vibrational states.

Starting with the theoretical line list, sorted by the final state term values, we searched for pairs and triplets of experimental lines that were within 1 cm$^{-1}$ of the predicted line positions, and formed combination differences within 0.005 cm$^{-1}$, where the final states were calculated using the precise term values for the states of v$_3$.[27, 62, 63] We compared the combination difference candidates to the Hamiltonian predictions, and assigned those displaying convincing agreement in wavenumber and relative intensities to predicted sets of combination differences. This yielded 66 assigned transitions, 1 of which had been previously observed and assigned as a 3LDR, confirming its final state. We could assign 3 other 4LDR lines by looking for combination differences between these and 3LDR transitions observed previously, again requiring agreement of 0.005 cm$^{-1}$ between the experimental final term values and reasonable agreement with predicted intensities. Further, each 4LDR line was assigned at most one predicted transition.

Table 1 gives a list of the thus proposed assignments, grouped into sets that share a final state. The list contains 76 assignments to 34 distinct final states between 8972 and 9461 cm$^{-1}$ with J-numbers between 2 and 9. In the vibrational assignment, a multiplicity index n$_v$ is explicitly introduced to distinguish different vibrational sub-states with the same symmetry for a given vibrational state. Eight of the reported lines were observed previously as 3LDR transitions and are flagged as such in the last column. They are included in the table since they form combination differences with 4LDR lines, and their transition wavenumber as well as the upper state term value of the combination difference were obtained from the 3LDR transitions. For the remaining combination differences, we calculated the upper term value as a weighted mean of the values from 4LDR lines forming it. The mean difference between assigned and predicted wavenumber of all reported lines is -0.070 cm$^{-1}$ and the standard deviation is 0.122 cm$^{-1}$. The standard deviation of the agreement in the experimental upper term values of the combination differences of the assigned 4LDR lines is 0.0013 cm$^{-1}$. This is more than the uncertainty originating from fits, pressure shift and FTS calibration; we attribute this to baseline issues that introduce systematic errors. Thus, we added the extra contribution of 0.0013 cm$^{-1}$ in quadrature to the line center uncertainty of all 4LDR lines in Table 1, and propagated this uncertainty to the upper term values.

Table 1. A list of transitions assigned in this work, grouped by the common final state, shown in Column 1. Column 2 shows the experimental transition wavenumbers, column 3 – the lower levels of the transitions from ExoMol[27]. Column 4 lists the transition wavenumbers from the Hamiltonian predictions[25], followed by the vibrational and rotational assignments of the upper and lower states, and their counting numbers. The last column flags if the line was observed as 4LDR or 3LDR.

| Upper state term value [cm$^{-1}$] | Experimental transition wavenumber [cm$^{-1}$] | Lower state term value [cm$^{-1}$] | Hamiltonian transition wavenumber [cm$^{-1}$] | V$_{up}$ | J$_{up}$ | C$_{up}$ | V$_{low}$ | J$_{low}$ | C$_{low}$ | flag |
|---|---|---|---|---|---|---|---|---|---|---|
| 8971.9732(9) | 5890.566(1) | 3081.406885(1) | 5890.393832 | 0 0 3 0 1F2 | 3A2 | 182 | 0 0 1 0 1F2 | 3A1 | 4 | 4LDR |
| 8971.9732(9) | 5920.300(1) | 3051.673419(1) | 5920.127586 | 0 0 3 0 1F2 | 3A2 | 182 | 0 0 1 0 1F2 | 2A1 | 4 | 4LDR |
| 8979.5366(9) | 5898.129(1) | 3081.406885(1) | 5898.153663 | 0 0 3 0 1F1 | 2A2 | 118 | 0 0 1 0 1F2 | 3A1 | 4 | 4LDR |
| 8979.5366(9) | 5927.864(1) | 3051.673419(1) | 5927.887416 | 0 0 3 0 1F1 | 2A2 | 118 | 0 0 1 0 1F2 | 2A1 | 4 | 4LDR |
| 9009.3793(8) | 5888.815(1) | 3120.565552(1) | 5888.979922 | 0 0 3 0 1F1 | 3A2 | 185 | 0 0 1 0 1F2 | 4A1 | 7 | 4LDR |
| 9009.3793(8) | 5927.971(1) | 3081.406885(1) | 5928.138712 | 0 0 3 0 1F1 | 3A2 | 185 | 0 0 1 0 1F2 | 3A1 | 4 | 4LDR |
| 9009.3793(8) | 5957.706(1) | 3051.673419(1) | 5957.872466 | 0 0 3 0 1F1 | 3A2 | 185 | 0 0 1 0 1F2 | 2A1 | 4 | 4LDR |
| 9043.7113(8) | 5923.144(1) | 3120.565552(1) | 5923.046682 | 1 4 0 0 1A1 | 3A2 | 189 | 0 0 1 0 1F2 | 4A1 | 7 | 4LDR |
| 9043.7113(8) | 5962.305(1) | 3081.406885(1) | 5962.205473 | 1 4 0 0 1A1 | 3A2 | 189 | 0 0 1 0 1F2 | 3A1 | 4 | 4LDR |
| 9043.7113(8) | 5992.039(1) | 3051.673419(1) | 5991.939226 | 1 4 0 0 1A1 | 3A2 | 189 | 0 0 1 0 1F2 | 2A1 | 4 | 4LDR |
| 9049.5261(9) | 5928.961(1) | 3120.565552(1) | 5929.033112 | 0 0 3 0 1F1 | 4A2 | 218 | 0 0 1 0 1F2 | 4A1 | 7 | 4LDR |
| 9049.5261(9) | 5968.119(1) | 3081.406885(1) | 5968.191902 | 0 0 3 0 1F1 | 4A2 | 218 | 0 0 1 0 1F2 | 3A1 | 4 | 4LDR |
| 9072.5375(8) | 5951.972(1) | 3120.565552(1) | 5952.019238 | 1 4 0 0 1A1 | 3A2 | 191 | 0 0 1 0 1F2 | 4A1 | 7 | 4LDR |
| 9072.5375(8) | 5991.131(1) | 3081.406885(1) | 5991.178028 | 1 4 0 0 1A1 | 3A2 | 191 | 0 0 1 0 1F2 | 3A1 | 4 | 4LDR |
| 9072.5375(8) | 6020.864(1) | 3051.673419(1) | 6020.911782 | 1 4 0 0 1A1 | 3A2 | 191 | 0 0 1 0 1F2 | 2A1 | 4 | 4LDR |
| 9072.8284(9) | 5950.346(1) | 3122.484494(1) | 5950.632403 | 0 2 2 0 2E | 4A1 | 240 | 0 0 1 0 1F2 | 4A2 | 5 | 4LDR |
| 9072.8284(9) | 5989.043(1) | 3083.784089(1) | 5989.331748 | 0 2 2 0 2E | 4A1 | 240 | 0 0 1 0 1F2 | 3A2 | 6 | 4LDR |
| 9087.2356(9) | 5915.165(1) | 3172.073005(1) | 5915.02497 | 1 4 0 0 1A1 | 4A1 | 241 | 0 0 1 0 1F2 | 5A2 | 7 | 4LDR |
| 9087.2356(9) | 5964.749(1) | 3122.484494(1) | 5964.614137 | 1 4 0 0 1A1 | 4A1 | 241 | 0 0 1 0 1F2 | 4A2 | 5 | 4LDR |
| 9100.4592(9) | 5928.386(1) | 3172.073005(1) | 5928.504226 | 0 0 3 0 1F1 | 5A1 | 266 | 0 0 1 0 1F2 | 5A2 | 7 | 4LDR |
| 9100.4592(9) | 5977.975(1) | 3122.484494(1) | 5978.093393 | 0 0 3 0 1F1 | 5A1 | 266 | 0 0 1 0 1F2 | 4A2 | 5 | 4LDR |
| 9103.0556(9) | 5866.571(1) | 3236.48512(2) | 5866.702839 | 0 0 3 0 1F1 | 5A2 | 283 | 0 0 1 0 1F2 | 6A1 | 9 | 4LDR |
| 9103.0556(9) | 5924.599(1) | 3178.45609(2) | 5924.731043 | 0 0 3 0 1F1 | 5A2 | 283 | 0 0 1 0 1F2 | 5A1 | 6 | 4LDR |
| 9103.3600(9) | 5865.949(1) | 3237.41210(2) | 5866.079398 | 0 0 3 0 1F1 | 5A1 | 267 | 0 0 1 0 1F2 | 6A2 | 8 | 4LDR |
| 9103.3600(9) | 5925.397(1) | 3177.96237(1) | 5925.528595 | 0 0 3 0 1F1 | 5A1 | 267 | 0 0 1 0 1F2 | 5A2 | 8 | 4LDR |
| 9106.173(1) | 6024.765(1) | 3081.406885(1) | 6024.77079 | 0 0 3 0 2F2 | 3A2 | 194 | 0 0 1 0 1F2 | 3A1 | 4 | 4LDR |
| 9106.173(1) | 6054.501(1) | 3051.673419(1) | 6054.50454 | 0 0 3 0 2F2 | 3A2 | 194 | 0 0 1 0 1F2 | 2A1 | 4 | 4LDR |
| 9108.3010(9) | 5985.818(1) | 3122.484494(1) | 5985.81871 | 0 0 3 0 2F2 | 3A1 | 174 | 0 0 1 0 1F2 | 4A2 | 5 | 4LDR |
| 9108.3010(9) | 6024.516(1) | 3083.784089(1) | 6024.51805 | 0 0 3 0 2F2 | 3A1 | 174 | 0 0 1 0 1F2 | 3A2 | 6 | 4LDR |
| 9152.2606(9) | 5974.299(1) | 3177.96237(1) | 5974.30042 | 0 0 3 0 2F2 | 4A1 | 248 | 0 0 1 0 1F2 | 5A2 | 8 | 4LDR |
| 9152.2606(9) | 6029.775(1) | 3122.484494(1) | 6029.77931 | 0 0 3 0 2F2 | 4A1 | 248 | 0 0 1 0 1F2 | 4A2 | 5 | 4LDR |
| 9161.1134(9) | 5855.164(1) | 3305.95049(2) | 5855.303495 | 0 0 3 0 1F1 | 6A2 | 318 | 0 0 1 0 1F2 | 7A1 | 8 | 4LDR |
| 9161.1134(9) | 5982.656(1) | 3178.45609(2) | 5982.797878 | 0 0 3 0 1F1 | 6A2 | 318 | 0 0 1 0 1F2 | 5A1 | 6 | 4LDR |
| 9190.7883(9) | 5953.376(1) | 3237.41210(2) | 5953.456661 | 1 4 0 0 1A1 | 6A1 | 340 | 0 0 1 0 1F2 | 6A2 | 8 | 4LDR |
| 9190.7883(9) | 6012.826(1) | 3177.96237(1) | 6012.905858 | 1 4 0 0 1A1 | 6A1 | 340 | 0 0 1 0 1F2 | 5A2 | 8 | 4LDR |
| 9192.60902(1) | 5882.95713(2) | 3309.65174(2) | 5883.112846 | 0 0 3 0 1F2 | 7A2 | 371 | 0 0 1 0 1F2 | 7A1 | 9 | 3LDR |
| 9192.60902(1) | 5951.34689(1) | 3241.26202(2) | 5951.502137 | 0 0 3 0 1F2 | 7A2 | 371 | 0 0 1 0 1F2 | 6A1 | 10 | 3LDR |
| 9192.60902(1) | 5956.123(1) | 3236.48512(2) | 5956.280166 | 0 0 3 0 1F2 | 7A2 | 371 | 0 0 1 0 1F2 | 6A1 | 9 | 4LDR |
| 9198.2148(9) | 6020.252(1) | 3177.96237(1) | 6020.328144 | 0 0 3 0 2F2 | 5A1 | 279 | 0 0 1 0 1F2 | 5A2 | 8 | 4LDR |
| 9198.2148(9) | 6075.731(1) | 3122.484494(1) | 6075.807033 | 0 0 3 0 2F2 | 5A1 | 279 | 0 0 1 0 1F2 | 4A2 | 5 | 4LDR |
| 9205.25170(1) | 5963.98956(1) | 3241.26202(2) | 5963.9899 | 0 0 3 0 2F2 | 5A2 | 297 | 0 0 1 0 1F2 | 6A1 | 10 | 3LDR |
| 9205.25170(1) | 6026.794(1) | 3178.45609(2) | 6026.79613 | 0 0 3 0 2F2 | 5A2 | 297 | 0 0 1 0 1F2 | 5A1 | 6 | 4LDR |
| 9209.7654(8) | 5903.990(1) | 3305.77733(2) | 5903.866756 | 0 0 3 0 1A1 | 6A1 | 342 | 0 0 1 0 1F2 | 7A2 | 10 | 4LDR |
| 9209.7654(8) | 5972.353(1) | 3237.41210(2) | 5972.231426 | 0 0 3 0 1 A1 | 6A1 | 342 | 0 0 1 0 1 F2 | 6A2 | 8 | 4LDR |
| 9209.7654(8) | 6031.802(1) | 3177.96237(1) | 6031.680623 | 0 0 3 0 1 A1 | 6A1 | 342 | 0 0 1 0 1 F2 | 5A2 | 8 | 4LDR |
| 9210.1640(9) | 5973.679(1) | 3236.48512(2) | 5973.653633 | 0 5 0 1 1 F1 | 6A2 | 326 | 0 0 1 0 1 F2 | 6A1 | 9 | 4LDR |
| 9210.1640(9) | 6031.708(1) | 3178.45609(2) | 6031.681838 | 0 5 0 1 1 F1 | 6A2 | 326 | 0 0 1 0 1 F2 | 5A1 | 6 | 4LDR |
| 9233.18249(1) | 5927.231(1) | 3305.95049(2) | 5927.44948 | 0 0 3 0 1 F1 | 7A2 | 383 | 0 0 1 0 1 F2 | 7A1 | 8 | 4LDR |
| 9233.18249(1) | 5996.697(1) | 3236.48512(2) | 5996.915658 | 0 0 3 0 1 F1 | 7A2 | 383 | 0 0 1 0 1 F2 | 6A1 | 9 | 4LDR |
| 9233.18249(1) | 5991.92035(1) | 3241.262016(1) | 5992.137628 | 0 0 3 0 0 F1 | 7A2 | 383 | 0 0 3 0 0 F1 | 7A2 | 10 | 3LDR |
| 9233.2525(9) | 5927.475(1) | 3305.77733(2) | 5927.686447 | 0 0 3 0 1 F1 | 7A1 | 365 | 0 0 1 0 1 F2 | 7A2 | 10 | 4LDR |
| 9233.2525(9) | 5995.841(1) | 3237.41210(2) | 5996.051117 | 0 0 3 0 1 F1 | 7A1 | 365 | 0 0 1 0 1 F2 | 6A2 | 8 | 4LDR |
| 9236.5327(9) | 5845.091(1) | 3391.44204(3) | 5845.312789 | 0 0 3 0 1 F1 | 7A1 | 366 | 0 0 1 0 1 F2 | 8A2 | 9 | 4LDR |
| 9236.5327(9) | 5921.551(1) | 3314.98214(2) | 5921.772275 | 0 0 3 0 1 F1 | 7A1 | 366 | 0 0 1 0 1 F2 | 7A2 | 11 | 4LDR |
| 9247.880(1) | 5932.901(1) | 3314.98214(2) | 5932.913902 | 0 2 2 0 2 F2 | 6A1 | 351 | 0 0 1 0 1 F2 | 7A2 | 11 | 4LDR |
| 9247.880(1) | 6075.805(1) | 3172.073005(1) | 6075.82354 | 0 2 2 0 2 F2 | 6A1 | 351 | 0 0 1 0 1 F2 | 5A2 | 7 | 4LDR |
| 9261.782(1) | 6024.369(1) | 3237.41210(2) | 6024.452395 | 0 0 3 0 2 F2 | 6A1 | 353 | 0 0 1 0 1 F2 | 6A2 | 8 | 4LDR |
| 9261.782(1) | 6083.822(2) | 3177.96237(1) | 6083.901592 | 0 0 3 0 2 F2 | 6A1 | 353 | 0 0 1 0 1 F2 | 5A2 | 8 | 4LDR |
| 9303.659(1) | 5988.678(1) | 3314.98214(2) | 5988.643121 | 0 2 2 0 1 F2 | 7A1 | 375 | 0 0 1 0 1 F2 | 7A2 | 11 | 4LDR |
| 9303.659(1) | 6066.245(1) | 3237.41210(2) | 6066.213839 | 0 2 2 0 1 F2 | 7A1 | 375 | 0 0 1 0 1 F2 | 6A2 | 8 | 4LDR |
| 9314.1987(9) | 5832.884(1) | 3481.31581(3) | 5833.077323 | 0 0 3 0 1 F1 | 8A1 | 431 | 0 0 1 0 1 F2 | 9A2 | 12 | 4LDR |
| 9314.1987(9) | 5999.216(1) | 3314.98214(2) | 5999.411075 | 0 0 3 0 1 F1 | 8A1 | 431 | 0 0 1 0 1 F2 | 7A2 | 11 | 4LDR |
| 9319.445(1) | 5832.766(2) | 3486.6780(5) | 5833.042554 | 0 0 3 0 1 F1 | 8A2 | 417 | 0 0 1 0 1 F2 | 9A1 | 11 | 4LDR |
| 9319.445(1) | 5921.873(1) | 3397.5730(4) | 5922.147334 | 0 0 3 0 1 F1 | 8A2 | 417 | 0 0 1 0 1 F2 | 8A1 | 12 | 4LDR |
| 9340.09129(2) | 5942.516(1) | 3397.5730(4) | 5942.51956 | 0 0 3 0 2 F2 | 7A2 | 404 | 0 0 1 0 1 F2 | 8A1 | 12 | 4LDR |
| 9340.09129(2) | 5951.75790(4) | 3388.33331(2) | 5951.75946 | 0 0 3 0 2 F2 | 7A2 | 404 | 0 0 1 0 1 F2 | 8A1 | 11 | 3LDR |
| 9340.09129(2) | 6030.43943(2) | 3309.65174(2) | 6030.44091 | 0 0 3 0 2 F2 | 7A2 | 404 | 0 0 1 0 1 F2 | 7A1 | 9 | 3LDR |
| 9340.09129(2) | 6098.82913(3) | 3241.26202(2) | 6098.8302 | 0 0 3 0 2 F2 | 7A2 | 404 | 0 0 1 0 1 F2 | 6A1 | 10 | 3LDR |
| 9348.13766(6) | 5950.563(2) | 3397.5730(4) | 5950.56566 | 0 2 2 0 1 E | 7A2 | 405 | 0 0 1 0 1 F2 | 8A1 | 12 | 4LDR |
| 9348.13766(6) | 6038.48580(6) | 3309.65174(2) | 6038.48701 | 0 2 2 0 1 E | 7A2 | 405 | 0 0 1 0 1 F2 | 7A1 | 9 | 3LDR |
| 9366.910(1) | 5885.594(1) | 3481.31581(3) | 5885.883241 | 0 2 2 0 2 E | 8A1 | 442 | 0 0 1 0 1 F2 | 9A2 | 12 | 4LDR |
| 9366.910(1) | 6051.927(1) | 3314.98214(2) | 6052.216994 | 0 2 2 0 2 E | 8A1 | 442 | 0 0 1 0 1 F2 | 7A2 | 11 | 4LDR |
| 9406.9399(9) | 5925.624(1) | 3481.31581(3) | 5925.870059 | 0 0 3 0 1 F1 | 9A1 | 463 | 0 0 1 0 1 F2 | 9A2 | 12 | 4LDR |
| 9406.9399(9) | 6015.498(1) | 3391.44204(3) | 6015.744325 | 0 0 3 0 1 F1 | 9A1 | 463 | 0 0 1 0 1 F2 | 8A2 | 9 | 4LDR |
| 9461.024(1) | 5974.989(2) | 3486.03403(3) | 5974.930586 | 0 2 2 0 2 E | 9A1 | 473 | 0 0 1 0 1 F2 | 9A2 | 13 | 4LDR |
| 9461.024(1) | 6063.118(1) | 3397.90702(3) | 6063.057343 | 0 2 2 0 2 E | 9A1 | 473 | 0 0 1 0 1 F2 | 8A2 | 10 | 4LDR |

## 6. Conclusions

We consider these proposed assignments as likely but not as definite as those assigned by 3LDR. They could, in principle, be tested by DR experiments using the pump transitions to populate the proposed lower states but that would be time consuming. One of the present authors has plans to further improve the accuracy of his Hamiltonian predictions by adjusting some parameters to fit the firmly assigned transitions to states of the P6 polyad. If the errors in the predictions can be substantially reduced (as has been the case for lower polyads), the assignments of the 4LDR transitions will be further improved. The experimental parameters of the present spectra were optimized for detection of 3LDR transitions; it is likely that addition of a buffer gas could be used to increase the steady state density of vibrationally excited states in the probe volume.

**Supplementary material**
The supplementary material contains a table with all transitions detected in this work.

**Acknowledgements**
This project is supported by the Knut and Alice Wallenberg Foundation (grant: KAW 2020.0303), and the Swedish Research Council (grant: 2020-00238). K.K.L. acknowledges funding from the U.S. National Science Foundation (grant: CHE-2108458) and the Wenner Gren Foundation (grant: GFOv2024-0010). M.R. acknowledges support from the French National Research Agency TEMMEX project (grant: 21-CE30-0053-01).

**Conflict of interest**
The authors declare no conflict of interest.

**Author declarations**
The authors have no conflicts to disclose.

Kevin K. Lehmann: Conceptualization (lead), Formal Analysis (equal), Methodology (equal), Validation (supporting), Writing/Original Draft Preparation (lead)

Isak Silander: Formal Analysis (equal), Methodology (equal), Data Curation (equal)

Adrian Hjältén: Formal Analysis (equal), Methodology (equal), Data Curation (equal), Validation (lead), Visualization (lead), Writing/Original Draft Preparation (supporting)

Michael Rey: Methodology (equal), Formal Analysis (supporting), Validation (supporting)

Aleksandra Foltynowicz: Funding Acquisition (lead), Project Administration (lead), Supervision (lead), Resources (lead), Writing/ Original Draft Preparation (supporting)

**Data availability statement**
The normalized interleaved spectra that support the findings of this study are available in the Zenodo database with the identifier https://doi.org/10.5281/zenodo.13341073.